# Thermal Management in Fine-Grained 3-D Integrated Circuits

Md Arif Iqbal, Naveen Kumar Macha, Wafi Danesh, Sehtab Hossain, Mostafizur Rahman
Computer Science Electrical Engineering, University of Missouri-Kansas City, MO, USA
mibn8@mail.umkc.edu, rahmanmo@umkc.edu

*Abstract*— **For beyond 2-D CMOS logic, various 3-D integration approaches specially transistor based 3-D integrations such as monolithic 3-D [1], Skybridge [2], SN3D [3] holds most promise. However, such 3D architectures within small form factor increase hotspots and demand careful consideration of thermal management at all levels of integration [4] as stacked transistors are detached from the substrate (i.e., heat sink). Traditional system level approaches such as liquid cooling [5], heat spreader [6], etc. are inadequate for transistor level 3-D integration and have huge cost overhead [7]. In this paper, we investigate the thermal profile for transistor level 3-D integration approaches through finite element based modeling. Additionally, we propose generic physical level heat management features for such transistor level 3-D integration and show their application through detailed thermal modeling and simulations. These features include a thermal junction and heat conducting nano pillar. The heat junction is a specialized junction to extract heat from a selected region in 3-D; it allows heat conduction without interference with the electrical activities of the circuit. In conjunction with the junction, our proposed thermal pillars enable heat dissipation through the substrate; these pillars are analogous to TSVs/Vias, but carry only heat. Such structures are generic and is applicable to any transistor level 3-D integration approaches. We perform 3-D finite element based analysis to capture both static and transient thermal behaviors of 3-D circuits, and show the effectiveness of heat management features. Our simulation results show that without any heat extraction feature, temperature for 3-D integrated circuits increased by almost 100K-200K. However, proposed heat extraction feature is very effective in heat management, reducing temperature from heated area by up to 53%.**

## I. Introduction

As 2-D CMOS scaling reaches its end [8], migration to 3-D provides possible pathways to continue scaling. However, 3-D IC approaches so far with die-die [9], wafer-wafer [10] and layer-layer [11] stacking retain 2-D CMOS's challenges and add new constraints like cost and manufacturing complexities, reliability degradation, and lack of thermal management [12]. In contrast to 2-D and 3-D CMOS based approaches, 3-D vertical device integration proves to be one of the most promising technologies for semiconductor industry that overcome the scaling challenges by stacking transistors while providing integrated solution for nanoscale device, circuit, connectivity [13]-[19]. Such vertical transistors pave way for a compact realization of arithmetic logic functions and depict unique interest for structured ASIC applications. 3-D vertical integration can be achieved by a number of ways ranging from monolithic 3-D IC to multi-lithic 3-D IC to Skybridge approach.

However, at nanoscale, due to lack of heat dissipation path and confined geometry 3-D vertical devices suffer greatly from self-heating issues leading to hot-spot development [20]. In particular, 3-D vertical nanowire transistors exhibit prominent thermal confinement [21]; longer travelling heat path to reach ambient because of the additional integration layers, stacking of transistors and insulating materials created during fabrication between layers also impedes heat dissipation. Although potentially significant heat dissipation through thermal vias [22], forced air-cooling [23], micro-fluidic based channel integration [24], metalized carbon-nanotubes and graphite nano composites [25] are addressed for various 3-D integration, thermal management remains a challenge for dense architectures with reliable, high performance function requirement targets [26].

So far, except Skybridge, no heat management approaches are present for transistor level 3-D integration. To address the thermal issue, we propose physical level heat extraction features thermal junction and thermal pillars. Thermal junctions are unique junctions that allow heat extraction from selected place in 3-D without interfering with the circuit functionality and thus enable flexibility to be placed any location of the 3-D structure. Connected to these junctions are thermal pillars, which are nano pillars/ TSVs/Vias that directly connect with ground and acts as heat dissipater.

In order to capture thermal profile of transistor-level 3-D integrated circuits and evaluate effectiveness of proposed heat extraction features, we have done circuit level thermal evaluation using Finite Element based Method (FEM). FEM based models are most accurate for thermal profiling of 3-D ICs because of its ability to handle complex geometries and nonhomogeneous material, greater efficiency and more flexibility [27][28][29]. We have expanded device level analysis to circuits, and applied circuit operating conditions and information about 3-D circuit layout and new heat extraction techniques in a unique way to determine thermal profile of 3-D vertical integrated circuits through FEM. Our model accounts for nanoscale material properties, nanoscale dimensions, heat flow path, 3-D circuit operation and layout. This model is capable of capturing both static and dynamic thermal behavior. Dynamic thermal analysis determines the temperature profile of an IC at any time [30].

Our FEM simulations indicate that temperature of a single vertical nanowire increases by 56K from ambient temperature (300K) during operating condition but when applied at logic implementation such as dynamic NAND gate, temperature increased by almost double (650K) and mostly at top transistor level. In fact, our simulation results suggest that for all emerging vertical 3-D integration approaches, top tier transistors contribute to more hot spot generation than the transistors located close to substrate. For example, without any heat extraction for M3D, temperature increases up to 450K from 300K and as in case of SN3D, temperature rise by almost 30K. However our results show that proposed heat extraction features are effective in reducing the temperature of a heated spot by 53% in the best case; the temperature can be brought down to normal operating temperature at 310K. These results are significant, and pave way for new circuit design paradigm with intrinsic thermal management.

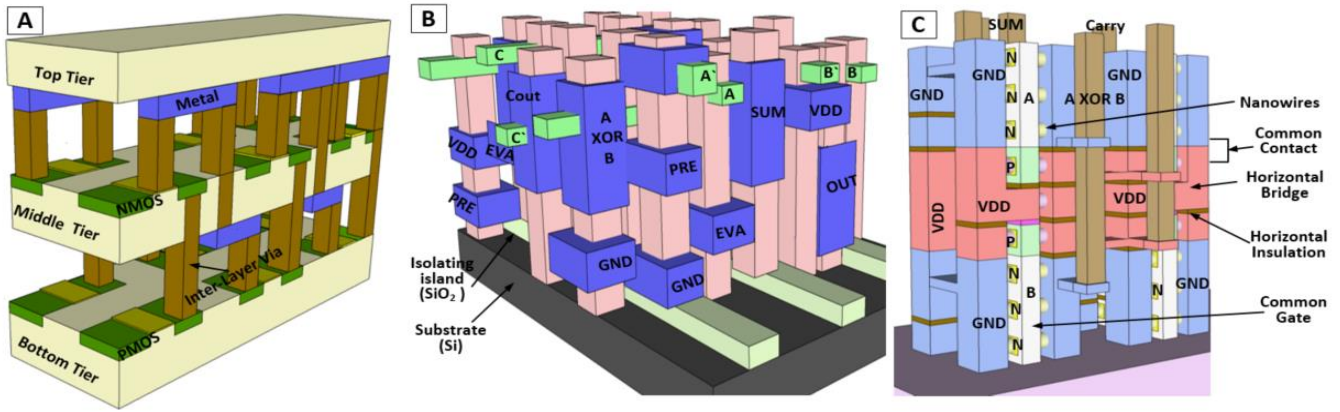

Fig. 1. Transistor-level 3-D integration approaches in literature [1][2][3]. A) Monolithic 3-D IC, B) Skybridge, C) SN3D

The remainder of the paper is organized as follows: Section II gives an overview about widely perused 3-D integration approaches. In section III, gives an overview about proposed heat extraction features. We present detail thermal modeling of 3-D integration in section IV and show the simulation results and effectiveness of proposed heat extraction features in Section V. Finally, the paper concludes in Section VI.

## II. EMERGING 3-D VERTICAL DEVICES : FABRIC OVERVIEW

Benefits of better immunity toward Short Channel Effect (SCE) and full compatibility with current planar technology [31] have made 3-D vertical device integration to be one of the most promising technology for the semiconductor industry. Among various 3-D vertical integration, monolithic 3-D IC [1] to multi-lithic 3-D IC [1] to SN3D [3] to Skybridge approach [2] achieves highest benefits. In the followings, we give brief overview of different 3-D integration approaches:

*(i) Monolithic 3-D[1]:*

Monolithic 3D IC is a vertical integration technology that builds up two or more tiers of devices sequentially. Meaning a substrate is completed first with devices then a second layer of devices is formed on top of the first one and so on. Therefore, sequentially one after another device layers are formed in fine-grained manners and they are connected with vias. Several monolithic 3-D integration processes have been proposed and several works on SRAM have also been proposed [32]-[38]. Such 3-D technology provides much higher benefits in terms of area, power performance than 2-D CMOS technology [1]. A typical monolithic 3-D IC is shown in Fig. 1A. In monolithic 3-D IC, we can place any 2-D standard cells and connect them through interlayer vias. As it can be illustrated from Fig. 1A that PMOS transistors are at the bottom and NMOS transistors are at the top layer and the two layers are interconnected through interlayer vias.

*(ii) Skybridge[2]:*

Skybridge a novel 3-D IC fabric technology where device, circuit, connectivity and manufacturing issues are co-addressed in a 3-D compatible manner [2]. This an example of another new direction for truly vertical 3-D IC.

Fig. 1B shows the architecture of Skybridge fabric. First, it starts with a regular array of uniform vertical nanowires that forms the Skybridge template. Second, its doping requirement is uniform, without regions, and done once at the wafer level. Finally, the various features of the fabric are realized through functionalizing this template with material deposition techniques. All inserted material structure features, regarding device, circuit style, connectivity, thermal management, are co-architected for 3-D compatibility, manufacturability, requirements and overall efficiency. It relies on six core fabric components to address 3-D device, circuit, connectivity, and manufacturing requirements. One of the major components is uniform vertical nanowires. Junctionless transistors that do not require abrupt doping variation within the device are active devices. These devices are stacked on nanowires and are interconnected using Bridges and Coaxial structures to implement logic and memory functionalities.

*(iii) SN3D[3]:*

A new radically different fabric concept for 3-D CMOS using stacked horizontal nanowires (SN3D) on a single die. It is a single die based 3-D CMOS concept using stacked horizontal nanowires.

Fig. 1C shows a typical example of SN3D fabric. In this approach, horizontally stacked suspended nanowires serve as building blocks or templates, and specially architected device, connectivity, insulation, and heat extraction features are formed onto these nanowires through material depositions for fabric assembly. Logic and memory implementation are through CMOS logic and fabric specific physical mapping scheme. High degree of connectivity and heat management in SN3D is achieved by utilizing fabric's intrinsic features. The connectivity and heat management approach in SN3D is in stark contrast to 3-D CMOS schemes, which are dependent on large-area TSVs and packaging-level heat management respectively. In addition, scalability in this fabric is determined by the ability to integrate more circuits vertically, and primarily rely on material deposition techniques that are more controllable in comparison to the extreme lithography dependent device-scaling paradigm of traditional CMOS. Prefabricated and doped stacked horizontal nanowires are building blocks and Gate-All-Around nanowire FETs (GAANWFETs), Common Gate (CG), Common Contact (CC), Horizontal Insulation (HI) and Horizontal bridges (HB) are architected features built on to these nanowires. N and p-type GAA-NWFETs are active devices. The CG feature allows multiple devices to be gated with single input in vertical and/or horizontal directions. CCs provide common contact for stacked GAA-NWFETs and allow degrees of freedom for connectivity in 3-D. HI provide isolation between adjacent Gate, Source and Drain contacts in horizontal direction and HBs provide connectivity between vertical stacks.

Though such 3-D integration provides integrated solution for nanoscale device, circuit and connectivity but such approach suffers sever thermal management issue. This issue is a critical challenge for these highly dense 3-D approaches because the stacked transistors are detached from the substrate i.e. heat sink, and transistor-level self-heating exacerbates the issue by leading to hot-spot development. From, Fig. 1A-C, it can be seen that for all cases top transistors are far away from the heat sink which increases thermal resistance path and eventually leads to high temperature at top transistors as we will also be showing in our simulation results. So far, there is no heat management approach present for vertical nanowire based transistor at device or circuit levels. To address such issue we propose a unique heat extraction feature in next section that is fabric centric and is applicable to any 3-D vertical devices due to flexibility to be optimized according to designer's choice.

### III. INTRINSIC THERMAL MANAGEMENT APPROACH: SPECIALIZED HEAT EXTRACTION PATH

Thermal management has become pivotal issue at nanoscale. As transistors are scaling down, heat dissipation paths are reducing, thus causing self-heating in transistors. The situation worsen when multiple transistors are stacked vertically and consequently, thermal resistance from heat source to heat sink increases. In this paper, we propose a thermal management approach that addresses such nanoscale thermal issues through specialized heat extraction path built in as core fabric component. Such approach is unique and different from traditional CMOS approaches where heat extractions are done at system level and also inadequate.

Proposed specialized heat extraction feature includes a heat conducting nano pillar (Fig. 2A-C) that carries heat from the heated region all the way to heat sink. To extract heat from logic nanowire, heat conducting nano pillar is connected with a terminal junction via metal line. So the fact that the thermal junction are put at those specific places (Fig. 2A-C), since its electrically insulated, it does not allow any conduction of current it just conducts heat. Then once the heat is extracted from that specific region, it dissipates heat towards the substrate via pillar and conductive bridge. All these features are intrinsic to the fabric and optimizable at physical fabric level for 3-D compatibility and can be implemented anywhere in the circuit without any loss of circuit functionality and performance.

The formation of Heat Conducting Nano Pillar is done with tungsten deposition whose thermal conductivity is 167 $Wm^{-1} K^{-1}$. The pillar is designed in such way to facilitate maximum heat dissipation having occupying minimum area. Nano Pillar is connected with a thermal junction for extracting heat from the heated region of nanowires. This thermal junction is made from deposition of $Al_2O_3$. Such oxide composite is electrically isolating which confirms logic implementation at nanowire remain unaltered. This junction gives the flexibility to be placed anywhere on nanowire and extract heat and dissipate through heat conducting nano pillar in conjunction with metal line. The metal line works as connecting bridge between conducting nano pillar and oxide terminal. Detailed dimensions & materials used for heat management features are given in Table I. In next section, we have discussed about modeling of heat conduction in 3-D vertical devices and given modeling approach of specialized heat extraction model.

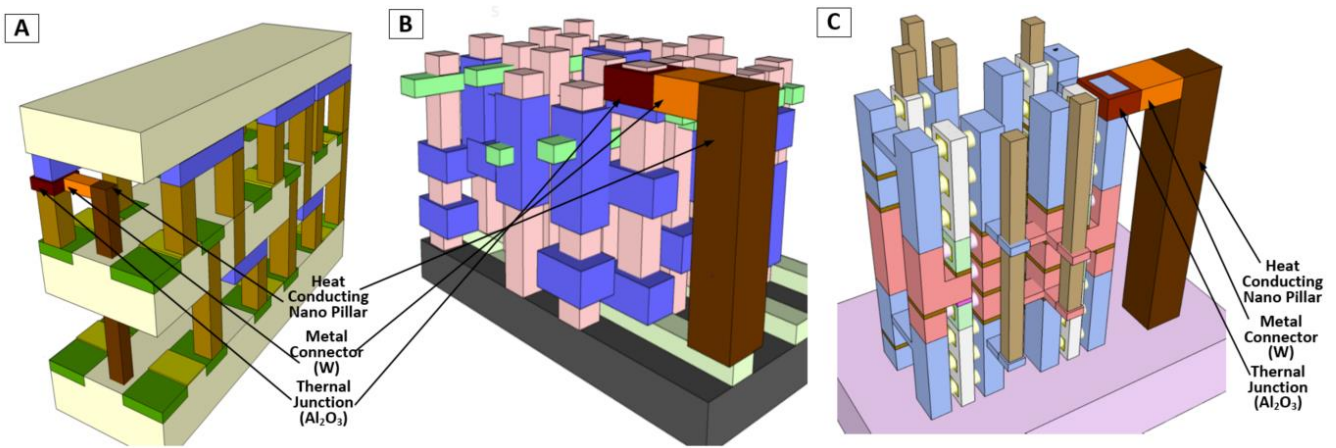

Fig. 2: Application of thermal extraction features at transistor-level 3-D integration. A-C) Customizable & adaptable heat conducting nano pillar connected with thermal junction to extract heat for Fig.1 3-D integration approaches.

TABLE I. DEVICE MATERIALS AND THEIR DIMENSIONS

| Skybridge | | | | SN3D | | | | Monolithic 3-D | | | |
| --- | --- | --- | --- | --- | --- | --- | --- | --- | --- | --- | --- |
| Region | Material | Dimension (LxWxT) nm | Thermal Conductivity Wm$^{-1}$K$^{-1}$ | Region | Material | Dimension (LxWxT)nm | Thermal Conductivity Wm$^{-1}$K$^{-1}$ | Region | Material | Dimension (LxWxT)nm | Thermal Conductivity Wm$^{-1}$K$^{-1}$ |
| Drain | Silicide | 10 x 16 x 16 | 45.9 [39] | Drain/Source Electrode | Ni | 24 x 32 x 34 | 90 [46] | Drain/Source Electrode | Ti | 24 x 24 x 32 | 21 |
| Drain/Source Electrode | Ti | 10 x 16 x 12 | 21 [40] | Channel | Doped Si | 16 x 16 x 16 | 13 | Channel | Doped Si | 16 x 32 x 32 | 13 |
| Channel | Doped Si | 16 x 16 x 16 | 13 [20] | Gate Oxide | HfO$_2$ | 16 x 22 x 3 | 0.52 | Source | Silicide | 24 x 24 x 32 | 45.9 |
| Source | Silicide | 10 x 16 x 16 | 45.9 | Gate Electrode | TiN | 16 x 32 x 34 | 1.9 | Gate Oxide | HfO$_2$ | 16 x 32 x 3 | 0.52 |
| Thermal Junction | AL$_2$O$_3$ | 210 x 60 x 16 | 30 [41] | Horizontal Insulation | Su-8 | 24 x 32 x 5 | 0.2 [47] | Gate Electrode | TiN | 16 x 32 x 32 | 1.9 |
| Gate Oxide | HfO$_2$ | 16 x 18 x 2 | 0.52 [42] | Thermal Junction | AL$_2$O$_3$ | 210 x 60 x 16 | 30 | Thermal Junction | AL$_2$O$_3$ | 210 x 60 x 16 | 30 |
| Gate Electrode | TiN | 10 x 16 x 6 | 1.9 [43] | Spacer | Si$_3$N$_4$ | 10 x 16 x 16 | 1.5 | Spacer | Si$_3$N$_4$ | 10 x 16 x 16 | 1.5 |
| Spacer | Si$_3$N$_4$ | 10 x 16 x 16 | 1.5 [44] | Metal Connector | W | 43.5x58x16 | 167 | Metal Connector | W | 43.5x58x16 | 167 |
| Heat Pillar | W | 168x36x36 | 167 [45] | Heat Pillar | W | 168x36x36 | 167 | Heat Pillar | W | 168x36x36 | 167 |

## IV. MODELLING OF HEAT CONDUCTION IN VERTICAL NANOWIRE

In order to, accurately determine thermal profile of 3-D vertical devices and quantify the effectiveness of proposed heat extraction feature, we have done both transistor & circuit level simulation using finite element based method (FEM). FEM simulations are known for their accuracy at atomic level precision. For thermal evaluation, we will use fine-grained model that accounts for thermal properties of materials, nanoscale dimensions, operating conditions, and 3-D layout. The methodology is illustrated in Fig. 3. In this methodology heat source of thermal simulations is calculated from transistor's electrical characteristics i.e. I-V characteristics and 3-D circuit's operating conditions. Through FEA, temperature values at the interfaces between interconnect and the top block and the bottom block are computed. This is followed by the mesh generation for represented structure and on each small granularity, we do self-consistent evaluations. This process, self-consistently calculate the heat distribution throughout the structure. Details of the model is explained below:

*(i) Fundamental Thermal PDEs:*

When an electrically conductive material is subjected to a current flow, Joule Heating occurs, which, in turns, causes increase in temperature in the material. The equation governing the PDEs is obtained as follows:

$$\rho Cp \frac{\partial T}{\partial t} + \nabla.(-k\nabla T) = Q \quad (1)$$

Where $\rho$ is the density, $Cp$ is the heat capacity, $k$ is the thermal conductivity and $Q$ denotes the heat generated due to electric potential applied at terminals. The generated heat $Q$ is related to current density and electrical field [48], the relation can be given as follows:

$$\nabla.J = Q \quad (2)$$

$$J = \left(\sigma + \epsilon_0\epsilon_r \frac{\partial}{\partial t}\right)E + J_0 \quad (3)$$

$$E = -\nabla.V \quad (4)$$

Here, $J$ is the current density, $\sigma$ is the electrical conductivity, $\epsilon_r$ is relative permittivity of material, $E$ is electric field and $V$ is electrical potential applied at the gate and drain terminal. The generated heat $Q$ flows through material modeled by using Fourier's law, which states that the conductive heat flux is proportional to the temperature gradient, and represented by

$$\nabla.(-k\nabla T) = Q \quad (5)$$

From equation (1), it can be observed that the variation of temperature is also a function of time from which we can calculate the dynamic behavior of the device. This allows the knowledge of heat profile at any circuit-operating period. For simulation purpose, we assumed 50% duty cycle of 100ns time. This time period is used electrical circuit evaluations in [2][3]. COMSOL-multiphysics, commercial finite element simulation tool [27], is used to solve these PDEs and capture the steady state and dynamic thermal behavior. For circuit level thermal analysis, we used NAND logic circuit configuration for earlier mentioned 3-D vertical devices.

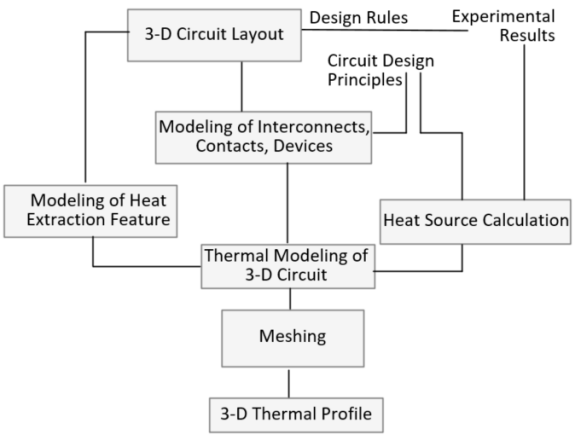

Fig. 3. Methodology for Finite Element Modeling

*(ii) Geometry Modeling:*

Details of geometry modeling will be taken from 3-D circuit layout. In this paper for modeling of Monolithic 3-D, we used design rules presented in [1]. We implemented NAND gate where two PMOS transistors are placed in the bottom stratum and two NMOS transistors are placed in top stratum. Size of inter-tier via is 50nm. Length of the ILV towards z-direction is 110nm. Total distance towards z-direction from the bottom substrate is 1.4um. Overall area for Monolithic 3-D is $183 \times 145 nm^2$ with channel length of 16nm. Local metal connection is of Aluminum. At the bottom of the substrate, a heat sink is placed with area of $600 \times 300 nm^2$. The geometry of SN3D fabric is taken form [3]. In this fabric, active devices are Gate-All-Around Junctionless nanowire FETs. Two n-type Junctionless nanowire FETs are placed at the bottom and two p-type Junctionless nanowire FETs are stacked on top to achieve NAND circuit configuration. A common metal connector is connected to the source side of the two p-type Junctionless transistors. Total area is $174 \times 114 nm^2$. A heat sink with same dimension as monolithic is placed at the bottom of the substrate. For modeling Skybridge circuit, we considered an n-type uniform V-GAA Junctionless transistor [2] and then implemented dynamic NAND gate, which follows the circuit style presented in [2]. Dynamic behavior of NAND gate is controlled by precharge (PRE), evaluate (EVA) control signals. In FEM based circuit simulations, we stacked all three transistors in single vertical nanowire. We assumed thermal conductivity of silicon nanowire to be 13 $Wm^{-1}K^{-1}$ since at nanoscale silicon conductivity reduces significantly [20]. Detail dimensions for all types of 3-D approaches are given in Table I. Proposed heat extraction paths were also placed according to fabric design principles outlined in Section III. The dimensions for heat extraction features mentioned in table I, can be modified according to fabric specifications. The pillars are directly connected to the substrate, which enables heat dissipation from hot spot region toward the heat sink.

*(iii) Grain Boundary Condition Modeling:*

For M3D, a simplified grain structure is used for the Inter-tier Vias and nearby wire connections. For simplicity, we used fewer numbers of local interconnects and TSVs which are only required for circuit implementation for our thermal analysis. However, in actual system, there will be more interconnect and TSVs. This grain structure was chosen to approximate the grain structure described in [1][2][3]. We first stack transistors blocks (for modeling purposes only) along the z direction sequentially on the top and the bottom of the substrate placed heat sink. For all three kinds' 3-D vertical circuit style, heat sources are assumed to be at the drain and gate blocks. As, Monolithic 3-D and SN3D follow combinational circuit styles, we applied voltages of 0.527V, 0.495V at the gate contact of the two pull-up transistors along with drain voltage to be at 0.8V. Since Skybridge follows dynamic circuit style, we assumed worst-case condition i.e. all the transistors are active. Gate voltages applied at Precharge, Input A, Evaluate transistors were 0.527V, 0.495V, 0.466V respectively. Drain voltage for each transistor was kept at 0.8V. All the voltages are calculated from devices electrical characteristics. The bottom surface of the heat sink is set to 300K (ambient). All other boundaries of the stack are set to be adiabatic.

For simplicity, a number of assumptions were made. We assume uniform grain and grain boundary geometry. Grain orientation or boundary propagation is not considered in this simulation. Moreover, quantum effects such were also ignores and thermal stress is negligible due to high thermal conductivity of proposed heat management features. We assume that grain and grain boundaries have the same initial atomic concentration and resistance.

## V. SIMULATION RESULTS

We have done extensive simulations to determine steady state and dynamic thermal profile of 3-D vertical transistor based circuits and to evaluate the effectiveness of intrinsic heat extraction features. In the following we present thermal simulations of vertical transistor based device and circuits.

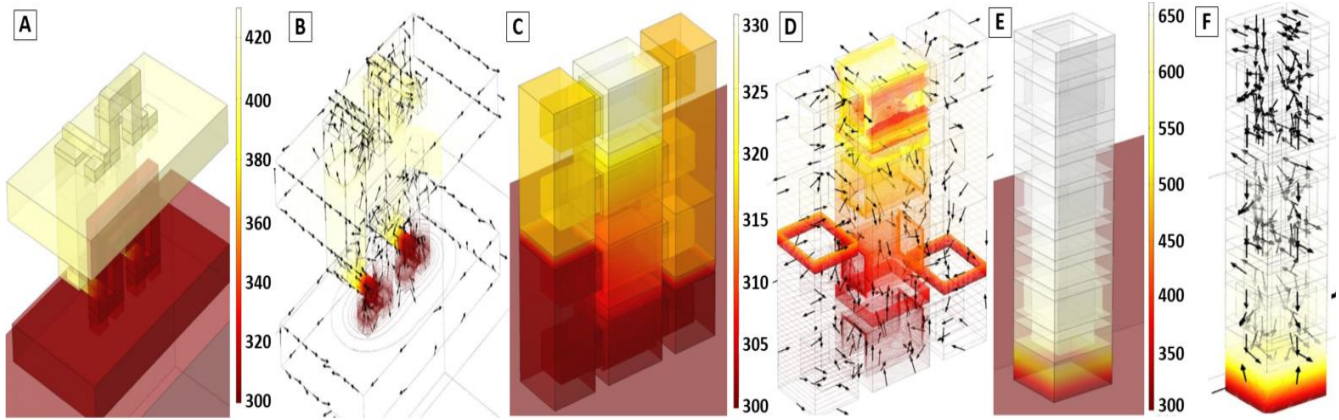

Fig. 4. Thermal profile of 3-D vertical Integrated Circuit. A) Monolithic 3-D. B) Heat flow path and temperature gradient of M3D. C) SN3D IC. D) Heat flow path and temperature gradient of SN3D. E) Skybridge D) Heat flow path and temperature gradient of Skybridge.

## A. Thermal Evaluation of 3-D Vertical transistors and Circuits

We emulated thermal behavior of different transistor–level vertical 3-D integration approaches without any heat extraction features. Fig 4 (A-F) shows the thermal profile and temperature gradient for M3D, SN3D and Skybridge. Temperature for M3D is much higher at the top tier than bottom tier. For top tier highest temperature is 420K, which is almost 120K rise in temperature. [49] also reported such rise in temperature for 3-D ICs. Lack of heat dissipation path, bulk material configuration and leakage current are the main causes of such high temperature. In addition, due to the sequential joining of layers in M3D, high-temperature processing on top layer affects the bottom layer; consequently, this yields to variability and reliability issues. From the temperature gradient (fig. 3B), it is seen that high heat at top tier is also contribution in hot spot creation at inter-layer vias. On the contrary, bottom tier being close to heat sink is maintaining ambient temperature. In addition, local interconnects for the bottom tier is at 300K. For Skybridge, highest temperature is at the top transistor to be 650K; almost 250K rise in temperature. Four vertical GAA nanowire transistors are stacked on top of each other. Top transistor is the furthest from the heat sink and lacks heat dissipation path. Heat has to follow the longest thermal resistance path. Moreover, for single vertical nanowire, in [50], it is reported that temperature can be increased by 60K due to only self-heating. Along with long thermal resistance, transistor-level self-heating exacerbates the issues by leading to such hot-spot development. As the heat flows towards the channel there are more heat dissipation path through drain/gate/source contact however source being closest to heat sink; temperature starts decreasing at source contact and reaches up to reference temperature (300K). For SN3D NAND logic circuit, temperature rise is only 30K. This rise in temperature is actually lower than other 3-D vertical approaches. This is mostly due the fabric style of SN3D. Highest temperature is at the gate region of top transistor due the confined geometry it lacks heat dissipation path. As in the case for drain contact (Fig. 4E), which is in for SN3D is also common contact for other transistor, heat is not localized due to sharing contact configuration. Moreover, source contact, which is also ground contact, is just placed below the drain contact separated by horizontal insulation. These give the benefits of extracting the additional heat created at the drain contact and dissipate towards substrate. However, from Fig. 1(A-F), it can be seen that for all cases top transistors are far away from the heat sink, which increases thermal resistance path and eventually leads to high temperature at top transistors.

## B. Thermal Behavior of Vertical 3-D Integrated Circuits Under Dielectric Medium

Usually active devices are wrapped with dielectric material. Additionally, during circuit operation, neighboring transistors who participate in logic implementation will give rise to an electric field. Any material experiencing such electric field will dissipate some of the electrical energy as heat results in creating dielectric medium. Such dielectric mediums are electrically insulated but thermally conductive. Therefore, dielectric medium in fact contributes in heat dissipation.

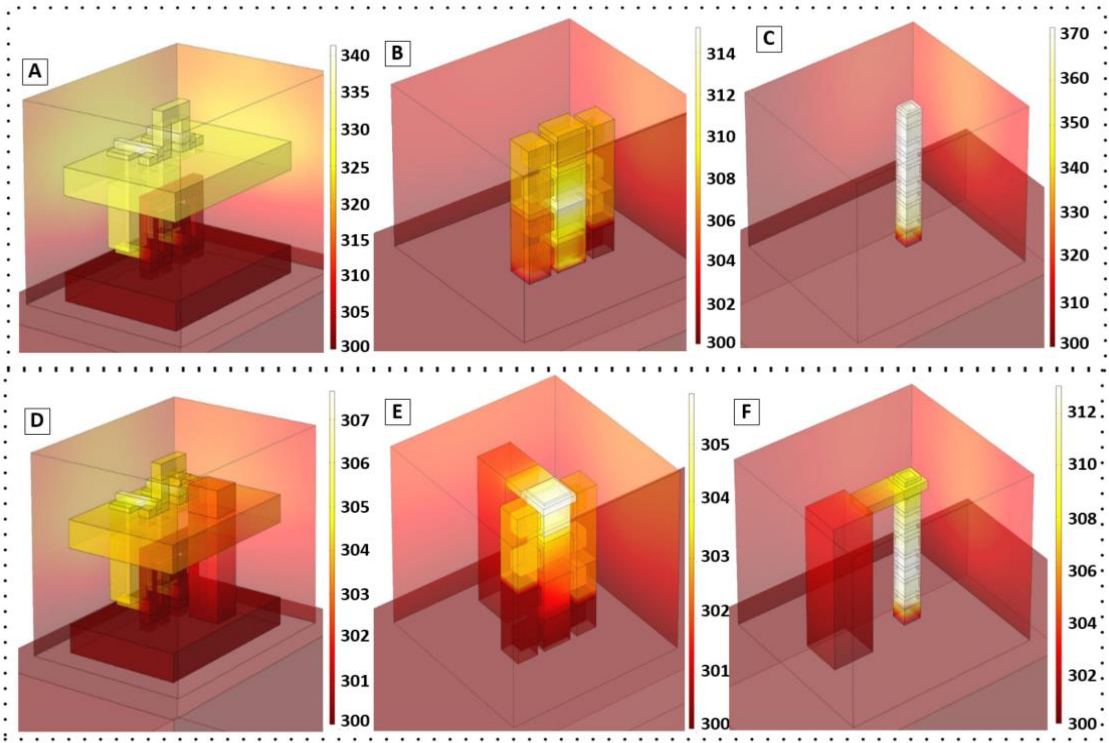

Fig. 5. Thermal behavior of 3-D vertical Integrated Circuit under dielectric medium and Effectiveness of fabric's intrinsic heat extraction feature. Dielectric medium contributing as heat dissipater for A) Monolithic 3-D. B) SN3D. C) Skybridge. D-F) Specialized heat extraction path reducing the temperature to 310K. by extracting additional heat from the most heated region and dissipating towards heat-sink for M3D, SN3D and Skybridge respectively.

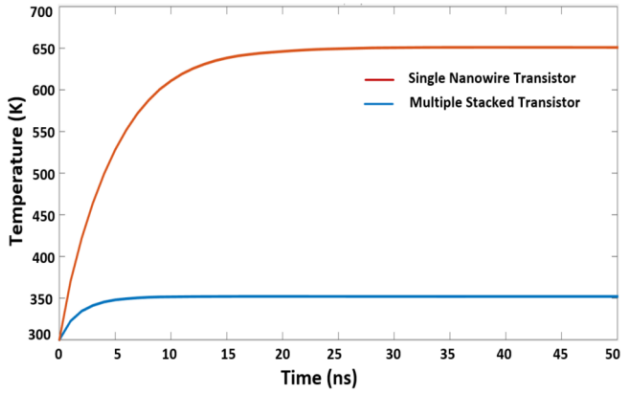

Fig. 6. Transient thermal behavior of vertically stacked Nanowire Transistor vs Single Nanowire Transistor

To capture the thermal behavior under dielectric medium we simulate heat propagation under dielectric medium for above-mentioned 3-D transistors. In our model, we approximate a rectangular box surrounding the nanowire circuit as dielectric medium as shown in Fig 5. As dielectric medium is electrically insulated, we assumed electrical conductivity to be 1 S/m. However, dielectric medium acts as heat dissipation path but has very low thermal conductivity of 0.3 $Wm^{-1}K^{-1}$[51]. From Fig. 5 (A-C), it can be seen that due to presence of dielectric medium temperature drops down on average 50K-200K, as for Skybridge, it reduces the temperature almost by 200K from 650K. Since the dielectric medium is widely covered around whole chip area, it can easily dissipate heat. However, reduced temperature is still not sufficient for normal circuit operation and may cause perturbation in logic implementation.

### C. Effectiveness of Intrinsic Heat Extraction Features for Thermal Management of Vertical 3-D Integrated Circuits

Effectiveness of proposed fabric's intrinsic heat extraction features is depicted in Fig. 5(D-E). Thermal junction is placed at the top transistor in conjunction with metal connector and heat conducting nano pillar to maximize heat dissipation from the most heated region of the circuit. Simulations are done considering circuit styles of different transistors and under same dielectric medium as described in previous section. Simulated results show that heat extraction features are effective in reducing the temperature for all the circuit styles especially transistor level 3-D integration that suffers most from heat management issues. The final temperature even for worst-case scenario is found to be around 310K, which is almost close to ambient temperature. For vertical nanowire transistors due to heat extraction features, it gets additional heat dissipation path that allows bringing down the temperature by almost 50%. This feature also has the flexibility to be placed anywhere in the circuit without any perturbation of circuit operation. As such, it allows placing the thermal junction even at the middle transistors, which will further bring down the temperature from 312K.

Above analysis gives us an idea about the steady state thermal profile of a circuit when the power source is applied. We also emulate dynamic thermal analysis of the circuits to get the temperature profile when the circuit is running. To capture the transient profile, 50% duty cycle is considered which is calculated from the device electrical stimulation. At first, we simulated for one single nanowire transistor and the expanded the simulation for whole circuit. Fig. 6 gives the comparison of dynamic temperature profile for single and multiple stacked transistors. For multiple stacked nanowire transistors, the temperature increases over 25ns of time and then stabilized. We further extended our dynamic thermal analysis on each of the transistor of different fabric. Fig. 7 gives a detail comparison of dynamic thermal profile for different fabric at top transistor. When the transistor are running without any heat extraction feature, temperature gradually increases. However, effectiveness of our proposed heat extraction can also be seen for fig.7. With heat extraction feature under dielectric medium, during any circuit operating time, the temperature always maintain the ambient temperature.

### VI. CONCLUSION

In this work, we showed thermal modeling (both steady state and dynamic) for emerging transistor level 3-D integration approaches. Our thermal modeling shows that without any heat extraction feature, temperature can increased by 100K-200K for vertical 3-D IC. To address the heat issue we showed a specialized heat management approach for 3-D vertical integrated circuit and proved its effectiveness through modeling and simulations. Simulation results show that our proposed heat extraction features can reduce the temperature of heated zone by 50% and can keep the temperature at ambient during circuit operation. Such heat management approach is unique and opens new opportunities for heat management of vertical 3-D transistors.

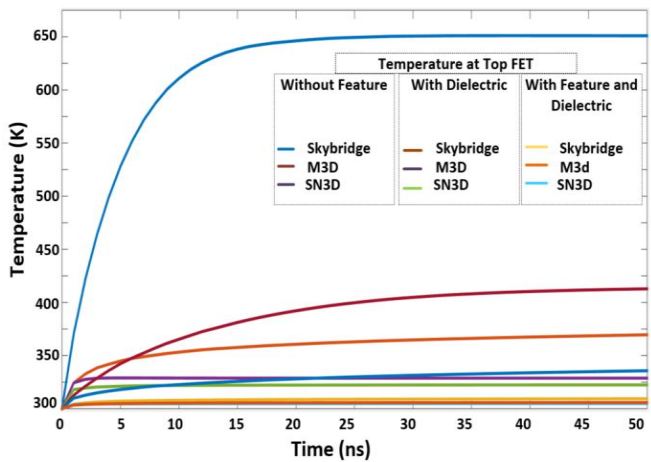

Fig. 7. Dynamic thermal behavior of top transistor of different 3-D IC under different condition.

### REFERENCES

1. Young-Joon Lee, et al., " Ultra-high density logic designs using transistor-level monolithic 3D integration," ICCAD, Nov. 2012
2. Rahman, M., Khasanvis, S., Shi, J., Li, M., & Moritz, "SkyBridge: 3-D Integrated Circuit Technology Alternative to CMOS." Preprint available at-http://arxiv.org/abs/1404.0.